\documentclass[conference]{IEEEtran}
\IEEEoverridecommandlockouts

\def\BibTeX{{\rm B\kern-.05em{\sc i\kern-.025em b}\kern-.08em
    T\kern-.1667em\lower.7ex\hbox{E}\kern-.125emX}}
    
\usepackage{cite}
\usepackage{lipsum}
\usepackage{tikz}
\usetikzlibrary{shapes,arrows}
\usepackage{standalone}
\usepackage{adjustbox}
\usepackage{amsmath,amssymb,amsfonts, mathtools}
\usepackage{algorithmic}
\usepackage{graphicx}
\usepackage{textcomp}
\usepackage{xcolor}
\usepackage{pgfplots}
\pgfplotsset{compat=1.14}
\usepackage{tabularx,booktabs}
\usepackage[font=small,labelfont=bf]{caption} % Required for specifying captions to tables and figures
\usepackage[subrefformat=parens]{subcaption}
\usepackage{pdfpages}
\newcolumntype{C}{>{\centering\arraybackslash}X}
\usepackage{multirow}
\definecolor{color_blue_light}{RGB}{166,206,227}
\definecolor{color_blue}{RGB}{31,120,180}
\definecolor{color_green_light}{RGB}{178,223,138}
\definecolor{color_green}{RGB}{51,160,44}
\definecolor{color_red_light}{RGB}{251,154,153}
\definecolor{color_red}{RGB}{227,26,28}
\definecolor{color_yellow}{RGB}{253,191,111}

\definecolor{color_red_red}{RGB}{228,26,28}
\definecolor{color_blue_blue}{RGB}{55,126,184}
\definecolor{color_green_green}{RGB}{77,175,74}
\definecolor{color_purple_purple}{RGB}{152,78,163}

\definecolor{color_green_blue}{RGB}{102,194,165}
\definecolor{color_orange}{RGB}{252,141,98}
\definecolor{color_gray}{RGB}{141,160,203}

\definecolor{google_green}{RGB}{50,185,100}
\definecolor{google_blue}{RGB}{0,87,231}
\definecolor{google_red}{RGB}{214,45,32}
\definecolor{google_yellow}{RGB}{255,167,0}

\definecolor{color1}{HTML}{ffb85a}
\definecolor{color2}{HTML}{ff655a}
\definecolor{color3}{HTML}{ff5ae9}
\definecolor{color4}{HTML}{705aff}

\begin{document}

\title{Model Fusion to Enhance the Clinical Acceptability of Long-Term Glucose Predictions\\
}

\author{\IEEEauthorblockN{Maxime De Bois}
\IEEEauthorblockA{Universit\'{e} Paris Saclay\\
CNRS-LIMSI\\
Orsay, France\\
maxime.debois@limsi.fr}
\and
% \IEEEauthorblockN{Moun\^{i}m A. El Yacoubi}
% \IEEEauthorblockA{Mines-Telecom Institute\\
% Teleecom SudParis\\
% \'{E}vry, France\\
% mounim.el\textunderscore yacoubi@telecom-sudparis.eu}
\IEEEauthorblockN{Moun\^{i}m A. El Yacoubi}
\IEEEauthorblockA{
T\'{e}l\'{e}com SudParis, Universit\'{e} Paris Saclay\\
SAMOVAR, CNRS\\
\'{E}vry, France\\
mounim.el\textunderscore yacoubi@telecom-sudparis.eu}
\and
\IEEEauthorblockN{Mehdi Ammi}
\IEEEauthorblockA{
Universit\'{e} Paris 8\\
Dept. of Computer Science\\
Saint-Denis, France\\
ammi@ai.univ-paris8.fr}}

\maketitle 

\begin{abstract}

This paper presents the Derivatives Combination Predictor (DCP), a novel model fusion algorithm for making long-term glucose predictions for diabetic people. First, using the history of glucose predictions made by several models, the future glucose variation at a given horizon is predicted. Then, by accumulating the past predicted variations starting from a known glucose value, the fused glucose prediction is computed. A new loss function is introduced to make the DCP model learn to react faster to changes in glucose variations.

The algorithm has been tested on 10 \textit{in-silico} type-1 diabetic children from the T1DMS software. Three initial predictors have been used: a Gaussian process regressor, a feed-forward neural network and an extreme learning machine model. The DCP and two other fusion algorithms have been evaluated at a prediction horizon of 120 minutes with the root-mean-squared error of the prediction, the root-mean-squared error of the predicted variation, and the continuous glucose-error grid analysis.

By making a successful trade-off between prediction accuracy and predicted-variation accuracy, the DCP, alongside with its specifically designed loss function, improves the clinical acceptability of the predictions, and therefore the safety of the model for diabetic people.

\begin{IEEEkeywords}
Glucose Prediction, Model Fusion, Clinical Acceptability, Artificial Neural Network
\end{IEEEkeywords}
\end{abstract}

\section{Introduction}

Diabetes, being the seventh leading cause of death in 2016, is one of the major diseases of the XXI century \cite{world2016global}. In order to avoid short-term (e.g., exhaustion, coma) or long-term (e.g., blindness, cardiovascular diseases) complications, diabetic people must maintain their blood glucose within acceptable ranges (i.e., between hypoglycemia and hyperglycemia). However, this task is far from easy given the high variety of factors influencing the variations of blood glucose (e.g., food intake, medications, physical activity, emotions).

Innovations that aim at helping diabetic people in their daily lives follow several leads. First, monitoring devices such as continuous glucose monitors (e.g., FreeStyle Libre \cite{olafsdottir2017clinical}) or medical coaching applications for diabetes (e.g., mySugr \cite{rose2013evaluating}) provide diabetic people with useful information such as current and past glucose values or calories intake history. Moreover, artificial pancreas start being commercialized and have already shown their effectiveness in managing the blood glucose of diabetic patients \cite{garg2017glucose}. Finally, nowadays, a lot of efforts have been focused towards the building of glucose predictive models \cite{oviedo2017review}. From the patient's past glucose, carbohydrate (CHO) intakes, and insulin infusions, those models try to predict future glucose values. 

In the past few years, a lot of different glucose predictive model architectures have been tried-out. Among them, Sun \textit{et al.} proposed a generic predictive model using Long Short-Term Memory (LSTM) and bidirectional LSTM neural networks to predict glucose at prediction horizons (PH) up to 60 minutes \cite{sun2018predicting}. De Paula \textit{et al.} studied the use of Gaussian Processes (GP) to predict future glucose values in an automated glucose controller based on reinforcement learning \cite{de2015controlling}. Besides, in their work, Georga \textit{et al.} analyzed different types of Extreme Learning Machine (ELM) networks for online short-term glucose prediction \cite{georga2015online}. Finally, Ben Ali \textit{et al.} proposed a tuning methodology for the architecture of Feed-Forward Neural Networks (FFNN) that aims at improving the glucose predictions of diabetic patients up to 60 minutes ahead \cite{ali2018continuous}.

Nonetheless, to this day, no algorithm is standing out from the others, with each of them having its own strengths and weaknesses \cite{debois2018study}. Several studies tried to make use of those specificities by combining the different predictions into a single one. Wang \textit{et al.} proposed the adaptive-weighted-average framework that combines glucose predictive models by weighing them based on their past errors \cite{wang2013novel}. Daskalaki \textit{et al.} built a hypoglycemia/hyperglycemia events warning system by combining autoregressive and recurrent neural networks models \cite{daskalaki2013early, botwey2014multi}. More recently, Jankovic \textit{et al.} studied a multi-step fusion methodology using ELM models for long-term glucose prediction \cite{jankovic2016deep}. Finaly, Yu \textit{et al.} proposed an adapative-filters-based fusion mechanism for short-term glucose prediction \cite{yu2018model}.

% bonne introduction. peut être insiscter plus sur les crtères d'évaluation en disant que les autres études ne les prennent pas en compte peut être ?

% autre remaque, il faut commencer par introduire la question scientifique et la controbution et les grandes lignes de ses avantages et son positionnement par rapprort à l'existant 

However, this particular question of long-term predictions remains an open problem. Due to the difficulty of the task, the models often output predictions that are inconsistent over time with a lot of high amplitude oscillations. This inconsistency directly impacts the clinical acceptability, measured by the Continuous Glucose-Error Grid Analysis (CG-EGA). To address this issue, we propose a novel model fusion algorithm, the Derivatives Combination Predictor (DCP), that bases its predictions on the prediction of glucose variations.

The paper is organized as follows. First, we introduce our algorithm. Then, we go through the different details about the  experiments we have conducted during the study. Finally, before concluding, we provide the reader with an analysis of the results.

\section{Derivatives Combination Predictor}

\subsection{Presentation of the Model}
% il me man que  la raison et le cheminament qui t'a ammané à cette prosotition. Tu peux t'appuyer sur l'état de l'art par ex. Plus généralement, il faut jsutifier les principaux choix techniques pris à chaque fois d'une façon ou d'une autre

\begin{figure*}
% \scalebox{0.5}{\includestandalone[width=\textwidth]{figures/DCP_schematics}}
% \includestandalone[width=\textwidth]{figures/DCP_schematics}
\centering
\includegraphics[width=\textwidth]{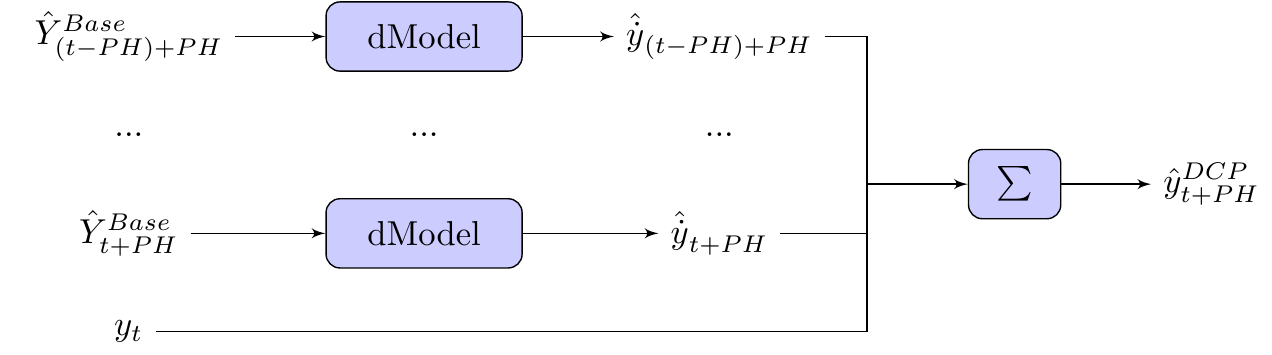}
\caption{DCP data flow, from the initial glucose predictions made by the \textit{Base} predictors we want to fuse to the prediction of the glucose derivatives by the dModel and the computation of the fused glucose predictions.}
\label{fig:dcp}
\end{figure*}

The goal of the DCP is to make the glucose predictions consistent with each other. In particular, it tries to make the difference between two consecutive predictions as close to the true glucose variation as possible. To do so, the DCP combines the predictions made by different predictors at a given prediction horizon (PH) into a single prediction following a two-step process (see Figure \ref{fig:dcp}).

First, at time $t$, the a model we call \textit{dModel} predicts the glucose variation at time $t+PH$, $\hat{\dot{y}}_{t+PH}$, from the past history of glucose predictions, $\hat{Y}^{Base}_{t+PH}$, made by the initial predictors $Base$.

% \subsubsection{$\Sigma$.}

Then, starting from the most recent glucose value known by the predictor (i.e., the glucose value at time $t$, when the prediction is made), the fused glucose prediction, $\hat{y}^{DCP}_{t+PH}$, is computed by accumulating the last PH predicted derivatives (see Equation \ref{eqn:derivativestovalue}).

\begin{equation}
\label{eqn:derivativestovalue}    
\Hat{y}_{t+PH}^{DCP} = {y}_t + \sum_{i=1}^{PH} \hat{\dot{y}}_{t+i}
\end{equation}

% \subsubsection{Example.} 
\textit{Example.}
We want to fuse two predictors, \textit{A} and \textit{B}, that forecast glucose values 120 minutes in the future (PH of 120 minutes). At every time-step \textit{t}, we forecast the glucose derivative (rate of change) 120 minutes in the future by giving the history of glucose predictions made by \textit{A} and \textit{B} to the dModel. With a history of 5, we give the dModel the predictions made by \textit{A} and \textit{B} at $t+120$, $t+119$, ..., $t+116$. 
The way the glucose variations are predicted depends on the nature of the dModel which can be any regression model (e.g., linear regressor or neural network). Once the glucose derivatives are predicted up to $t+120$, we can compute the glucose prediction at $t+120$ by using the last 120 glucose derivative predictions and the current glucose value using Equation \ref{eqn:derivativestovalue}.

\subsection{Learning of the dModel} \label{subsec:dModelLearning}

% La par ex, tu parles de FFNN. Tu peux faire reference à des travaux pour les propriété que tu évoques : flexibility & ability to model complex non-linear functions. Meme remaque pour le reste des arguments

While any supervised regression model can fit into the dModel, we chose to use a FFNN for its flexibility and its ability to model complex non-linear functions. To enhance the accuracy of the glucose predictions computed from the predicted derivatives, we introduce a new loss function to be used inside the FFNN-based dModel : the Derivatives-Biased Mean-Squared Error (MSE\textsubscript{DB}).

The MSE\textsubscript{DB} (see Equation \ref{eqn:loss}) is quite similar to the MSE loss function. For every sample $i$, with a small enough value of $\sigma$, the squared error will be scaled by $1 \pm \gamma$ depending on the value of $\hat{\dot{y}}_i / \dot{y}_i $. If $\hat{\dot{y}}_i / \dot{y}_i > 1$, meaning either $\hat{\dot{y}}_i > \dot{y}_i > 0$ or $\hat{\dot{y}}_i < \dot{y}_i < 0$, then the loss attributed to this sample is scaled down; if not, it is scaled up.

\begin{equation}
    MSE_{DB}(\boldsymbol{\dot{y}},\boldsymbol{\hat{\dot{y}}})=\frac{1}{n} \sum_{i=1}^n (1+\gamma \cdot \tanh(\frac{1-{\hat{\dot{y}}_i}/{\dot{y}_i}}{\sigma}))(\dot{y}_i-\hat{\dot{y}}_i)^2
\label{eqn:loss}    
\end{equation}

% meme remarque pour le dernier argument : this make our model react faster .... 

Intuitively, using this loss function during training means that we encourage the model to predict derivatives with the same sign as and higher absolute values than the true values. In practice, this makes our model react faster to changes in variations of glucose, and, therefore, make the predictions more accurate.

\begin{figure*}
\centering
\begin{adjustbox}{width=0.75\textwidth}
% \includestandalone[width=\columnwidth]{figures/loss_graphs}
\includegraphics[width=\textwidth]{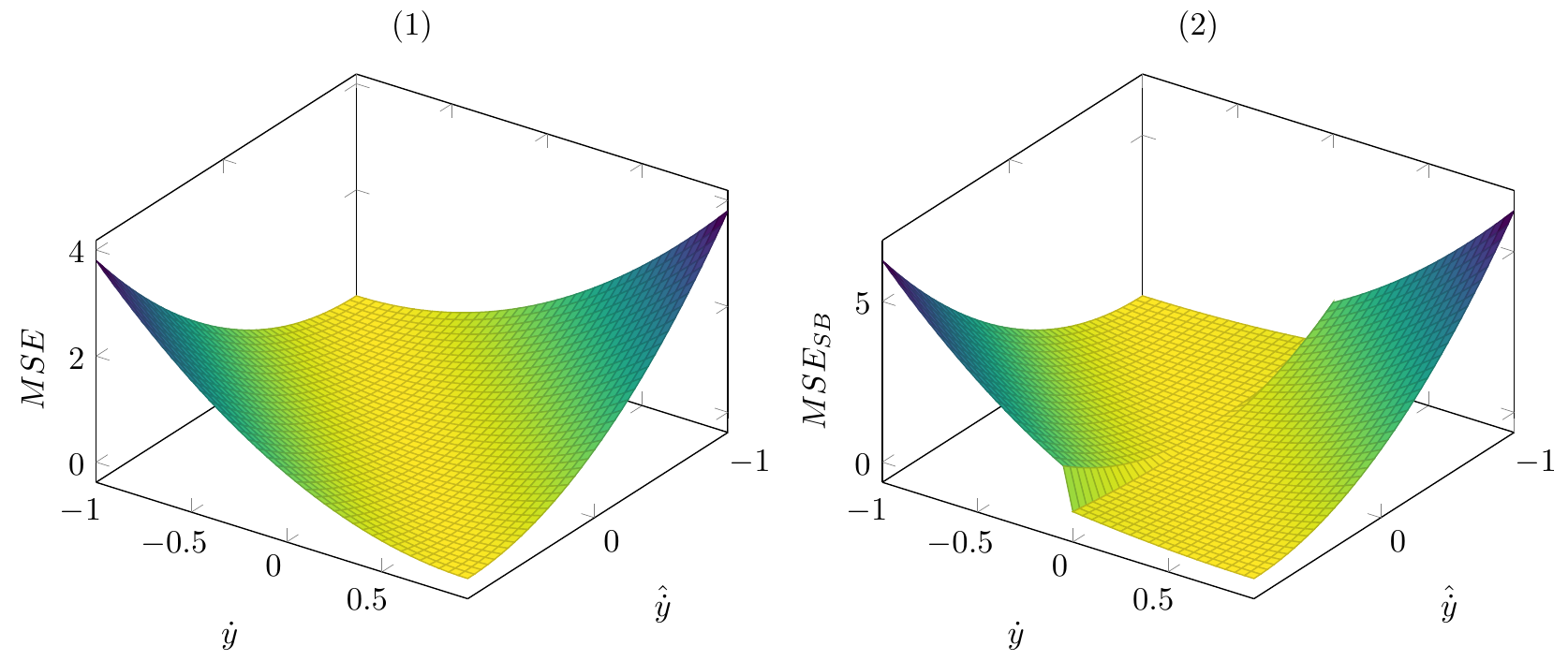}
\end{adjustbox}
\caption{Surface plots of the per sample $MSE$ (1) and $MSE_{DB}$ with $\gamma=0.65$ and $\sigma=10^{-5}$ (2) loss functions.}
\label{fig:graph}
% \end{center}
\end{figure*}

\section{Experimental Results}

\subsection{Experimental Data}

%c'est bien, mais il faut le trouner differement. 
%il faut dire que pour valider ton approche, on doit s'appuyer sur les données d'un simulateur approuvé par la FDA .... et que c'est appyé par la litrature dans le domaine (comme tu le dis à la fin). il faut dire que le simualteur genere des données réaliste de meme nature que celle générées par des lecture de glycémie, etc. en résumé dire que c'est trés proche de la réalité.

In this study, we used the 10 \textit{in-silico} children population from the UVA/Padova Type 1 Diabetes Metabolic Simulator (T1DMS) \cite{man2014uva}. T1DMS has been approved by the Food and Drug Administration in the United States as a substitute to clinical testing and is, therefore, extensively used in the glucose prediction literature \cite{oviedo2017review}. Compared to data coming from real patients, which are sensitive and oftentimes impossible to be shared, simulated data can be reproduced. Therefore, using the simulator makes our results reproducible and available for comparison.

The simulation lasted 28 days and outputted three different time-series sampled every minute: glucose value, insulin infusion and carbohydrate (CHO) intake over time. To account for the diversity of real-life situations, the subjects have been put under the following daily open-loop scenario:

\subsubsection{Meals} Every day is made of 3 meals, with each meal's timing and CHO amount being sampled from Gaussian distributions. The timing Gaussian distributions have a variance of 0.5 and a mean of 7h, 13h and 20h respectively. The CHO amount Gaussian distributions have a variance of 0.5 and a mean of 40g, 85g and 60g respectively. Every meal lasts 15 minutes.

\subsubsection{Insulin Boluses} At the start of every meal an insulin bolus is taken. The value of the bolus is sampled uniformly between 0.7 and 1.3 times the optimal insulin bolus. The optimal insulin bolus is computed by the simulator from the child's personal carbohydrate-to-insulin ratio.

\subsubsection{Basal Insulin} The basal insulin, computed by the simulator, is constant and optimal for every child.

% \textcolor{red}{
% The predictions of three different predictors\footnote{Other models have been tried out in addition to those three, but, as they did not show any further improvement to the results, we did not keep them in the final study.} have been used as the input to the DCP: a feedforward neural network (FFNN), a Gaussian process regressor (GP), and a extreme learning machine network (ELM). Those models have been personalized and trained on 10 virtual type-1 diabetic children from the T1DMS software (which is approved by the Food and Drug Administration in the United States) to predict future glucose values at a PH of 120 minutes.}

% \textcolor{red}{For every child, we have 28 different days worth of data, with 1440 predictions per day (one prediction every minute). The dataset has been split into train/test subsets to use a 14-fold cross-validation (the sequential nature of each day has been kept intact to make the subsets consistent). Every train/test split has been standardized (zero-mean and unit-variance).}

\subsection{Base Predictors} 

% \footnote{Other models have been tried out in addition to those three, but, as they did not show any further improvement to the results, we did not keep them in the final study.}

The predictions of three popular (in the context of glucose prediction) predictors have been used as the input to the DCP: a Feed-forward Neural Network (FFNN), a Gaussian Process regressor (GP), and an Extreme Learning Machine network (ELM) \cite{oviedo2017review,ali2018continuous,de2015controlling,georga2015online,jankovic2016deep,debois2018study}. All the models have been optimized through the tuning of their hyperparameters.

\subsubsection{Data Preprocessing} Keeping their sequential nature intact, the time-series outputted by the simulator have been grouped by days and then split into train and test subsets. The splitting has been done according to a 4 fold cross-validation used during the training and testing of the models. For each testing fold, a third of the training set constitutes the evaluation set, used to tune the models. After data normalization (zero-mean and unit-variance), the time-series are fed to the models as histories of the past 60 values (1 hour-long history).

\subsubsection{GP} The GP model has been implemented with a dot-product kernel \cite{debois2018study}. Whereas the kernel coefficient and the inhomogeneity have been set to $0.5$ and $0.01$ respectively, the noise-controlling hyperparameter has been grid-searched inside the $[10^{-2}, 10^{1}]$ range.

\subsubsection{ELM} Given the inherent simplicity of ELM models in general (no training and close to no tuning are needed), we optimized the number of neurons inside its single hidden layer within the $[1, 20160]$ range (20160 being the number of training samples) \cite{huang2006extreme}. To reduce the overfitting of the model to the training set, we added a L2 penalty ($10^{1}$) to the weights of the neurons.

\subsubsection{FFNN} The FFNN model is made of 4 hidden layers of respectively 128, 64, 32, and 16 neurons. We used the scaled exponential linear unit (SELU) activation function \cite{klambauer2017self}. The model is trained by the Adam optimizer with the mean-squared error (MSE) loss function, mini-batches of size $1500$, initial learning rate and decay of $10^{-3}$ and $10^{-4}$ respectively. To fight the overfitting of the model, we used several regularization methods, namely a L2 penalty of $10^{-4}$ and early stopping.

\subsection{DCP Implementation}

The FFNN-based dModel is made of 3 hidden layers of respectively 256, 128, and 64 neurons (ReLU activation function). The network takes as input the history of the 10 past predictions from the 3 base predictors, making-up to 30 inputs. 

For every train/test split, the network has been trained for 500 epochs with mini-batches of size 2500. To avoid the overfitting of the network to the training set, we used batch normalization layers (applied between the outputs of the neurons and the activation functions) and dropout (rate of 0.5). Furthermore, early stopping has been applied to the root-mean-squared error of the glucose predictions (computed with Equation \ref{eqn:derivativestovalue}) made on the evaluation set ($1/6$ of the training set, not used during training). Finally, the $\gamma$ and $\sigma$ coefficients of the $MSE_{DB}$ loss function has been optimized through grid-search, ending with a value of 0.65 and $10^{-5}$ respectively.

\subsection{Fusion Models for Comparison} 

% pourquoi ces deux modèles ? arguements ? refs ?

In order to evaluate the performance of the DCP, we implemented two other fusion algorithms: a model we call the Artificial neural network Combination Predictor (ACP) and the Adaptive-Weighed-Average (AWA) fusion algorithm from Wang \textit{et al.} \cite{wang2013novel}.

\subsubsection{ACP} The ACP model is a FFNN with the exact same architecture as the FFNN-based dModel. However, instead of predicting the glucose derivatives, it directly predicts the future glucose values. Therefore, the MSE\textsubscript{DB} loss function cannot be used since it is tailored to the dModel. The traditional MSE loss function has been used in its stead. The purpose of this model is to compare the DCP, a fusion model that predicts the future glucose values through the prediction of the future variations, to a model that directly predicts the future glucose values.

\subsubsection{AWA} The idea behind the AWA model is that every base predictor is assigned a weight based on its past recent errors. The weights are dynamically changed with the knowledge of new prediction errors. To strenghen the impact of the most recent errors on the weights, a forgetting factor is used \cite{wang2013novel}. It has been optimized through grid search and ended up with a value of $0.99$.

\subsection{Evaluation Metrics}

% paprler sur la compmplémentarité et le coté clinique et medical deja ici. meme si tu le développes encore à la fin.

In this study, we use three complementary metrics to evaluate the models: the Root-Mean-Squared Error of the prediction (RMSE), the Root-Mean-Squared Error of predicted variations (dRMSE), and the Continuous Glucose-Error Grid Analysis (CG-EGA). Whereas the RMSE and the dRMSE provide with the accuracy of the predictions, the CG-EGA measures the clinical acceptability of the models.

\subsubsection{RMSE}

The RMSE (see Equation \ref{eqn:rmse}, with $\boldsymbol{y}$ and $\boldsymbol{\hat{y}}$ being, respectively, the true and predicted glucose values) is a standard metric to evaluate regression models and in particular glucose predictive models. It provides a measure of the average accuracy of the glucose predictions. 

\begin{equation}
    RMSE(\boldsymbol{y},\boldsymbol{\hat{y}}) = \sqrt{\frac{1}{n}\sum_{i=1}^n(y_i - \hat{y}_i)^2}
    \label{eqn:rmse}
\end{equation}

\subsubsection{dRMSE} The dRMSE is simply the RMSE applied to the derivatives of the glucose predictions instead of the glucose predictions themselves. It gives a measure of the accuracy of the variations of the predictions when compared to the true variations.

\begin{equation}
    dRMSE(\boldsymbol{y},\boldsymbol{\hat{y}}) = \sqrt{\frac{1}{n-1}\sum_{i=1}^{n-1}(\Delta y_i - \Delta\hat{y}_i)^2}
    \label{eqn:drmse}
\end{equation}

% \textcolor{red}{The ESOD measures the energy of the acceleration of the predicted glucose signal (see Equation \ref{eqn:psod}) relative to the true signal. In other words, it measures how much the predictions oscillates and therefore the stability of the model over time.}

% \textcolor{red}{\begin{equation}
%     % PSOD(\boldsymbol{y},\boldsymbol{\hat{y}}) = {\frac{1}{n}\sum_{i=1}^n(\Delta^2y_i - \Delta^2\hat{y}_i)^2}
%     % PSOD(\boldsymbol{\hat{y}}) = {\frac{1}{n}\sum_{i=1}^n(\Delta^2\hat{y}_i)^2}
%     \label{eqn:psod}
%     ESOD(\boldsymbol{y},\boldsymbol{\hat{y}}) = \frac{\sum_{i=1}^n(\Delta^2\hat{y}_i)^2}{\sum_{i=1}^n(\Delta^2y_i)^2}
% \end{equation}}

\subsubsection{CG-EGA}

The CG-EGA is the most used metric in evaluating glucose predictive models as it measures the clinical acceptability of the predictions \cite{oviedo2017review}. In particular, it assesses, for every prediction, depending on the glycemia region (hypoglycemia, euglycemia\footnote{The euglycemia region is the region between hypoglycemia and hyperglycemia (between 70 $mg/dL$ and 180 $mg/dL$).}, hyperglycemia), the dangerousness of making such a prediction. This is very useful in the context of diabetes management since prediction errors can threaten the life of the patient.

Technically, the CG-EGA is made of two evaluation grids: the point-error grid analysis (P-EGA) and the rate-error grid analysis (R-EGA). While the P-EGA determines the clinical acceptability of the predictions themselves, the R-EGA focuses on the rates of change (the difference between two consecutive glucose predictions). The clinical acceptability is described by grades, from A to E, for both grids (Figure \ref{fig:cg-ega} provides the reader with a graphical example of the two grids). The overall clinical acceptability of the prediction is assessed by combining the two grids and classifying the prediction as either an accurate prediction (AP), a benign error (BE) or an erroneous prediction (EP). If a prediction and its associated derivative are both classified into the A or B categories, the prediction is then an AP.

\subsubsection{Complementarity of the Chosen Metrics} Whereas the RMSE and the P-EGA evaluate the accuracy of glucose predictions, the dRMSE and the R-EGA focus on the predicted glucose rates of change. We can then say that, generally, improving the RMSE  improves the P-EGA, and, improving the dRMSE improves the R-EGA. 

The CG-EGA is the most important metric as it determines if a predictive model is safe to use by diabetic people. While being very self-explanatory (in its simplified representation), it is a very complex metric. This implies that humans can have a hard time comparing models solely using the CG-EGA as the models might have different strengths and weaknesses.
In the other hand, the RMSE and the dRMSE, being single value metrics, are very simple. This makes the comparison between models fast and straightforward.

% \textcolor{cyan}{explain relationship between RMSE, dRMSE and CG-EGA and their complementary nature}

\subsection{Results}

The results of our study are represented by Table \ref{table:general_results}. In this table, the performances of the models, in terms of RMSE (in $mg^2/dL^2$), dRMSE (in $mg^2/dL^2/s$), and CG-EGA (in percentage of predictions falling into the different categories), are given. 

The difference between DCP 1 and DCP 2 relies on the loss function they use: DCP 1 uses the traditional MSE and DCP 2 uses the MSE\textsubscript{DB}.

% \textcolor{red}{The GP, ELM and FFNN models are the initial predictors used  for the fusion: they constitute our baseline. The AWA model implements the Adaptive-Weighed-Average framework from Wang \textit{et al.} (forgetting factor of 0.99) \cite{wang2013novel}. The ACP model is a simple FFNN that has the same architecture as the FFNN-based dModel, but predicts directly the future glucose values (instead of the derivatives). Finally, DCP \#1 and DCP \#2 are DCP models that use, respectively, the $MSE$ and $MSE_{DB}$ loss functions during their respective training.}

% \textcolor{cyan}{provide explanation concerning DCP 1 vs DCP 2}

\begin{table*}
    \caption{Performances of the models with mean $\pm$ standard deviation across the children population.}
    \label{table:general_results}
    \begin{tabularx}{\linewidth}{l||C|C|C|C|C}
        \toprule
        \multirow{3}{*}{\textbf{Models}} &  \multirow{3}{*}{\textbf{RMSE}} & \multirow{3}{*}{\textbf{dRMSE}} &  \multicolumn{3}{c}{\textbf{CG-EGA}} \\
        & & & Accurate & Benign & Erroneous \\
        & & & Predictions & Errors & Predictions \\
         \midrule
         \textbf{GP} & \textbf{48.47} {$\pm$} 10.02 & \textbf{5.12} $\pm$ 1.14 &  \textbf{68.31} $\pm$ 7.13 & \textbf{19.94} $\pm$ 4.88 & \textbf{11.75} $\pm$ 3.40 \\
         \textbf{ELM} & \textbf{42.30} $\pm$ 7.13 & \textbf{8.76} $\pm$ 1.33 & \textbf{73.62} $\pm$ 3.45 & \textbf{20.32} $\pm$ 2.11 & \textbf{6.05} $\pm$ 1.86\\
         \textbf{FFNN} & \textbf{37.79} $\pm$ 6.11 & \textbf{7.64} $\pm$ 0.79 & \textbf{73.99} $\pm$ 3.14 & \textbf{21.06} $\pm$ 1.99 & \textbf{4.95} $\pm$ 1.80 \\
         \midrule
         \textbf{ACP} & \textbf{36.14} $\pm$ 6.48 & \textbf{3.50} $\pm$ 0.92 & \textbf{78.10} $\pm$ 3.74 & \textbf{16.83} $\pm$ 2.38 & \textbf{5.07} $\pm$ 1.77\\
         \textbf{AWA} & \textbf{37.51} $\pm$ 6.35 & \textbf{5.29} $\pm$ 0.71 & \textbf{76.92} $\pm$ 2.79 & \textbf{17.61} $\pm$ 1.61 & \textbf{5.46} $\pm$ 1.50\\
         \textbf{DCP 1} & \textbf{52.68} $\pm$ 7.65 & \textbf{1.32} $\pm$ 0.28 & \textbf{84.52} $\pm$ 3.40 & \textbf{7.55} $\pm$ 2.11 & \textbf{7.93} $\pm$ 1.70\\
         \textbf{DCP 2} & \textbf{49.37} $\pm$ 8.61 & \textbf{1.33} $\pm$ 0.29 & \textbf{85.53} $\pm$ 4.43 & \textbf{7.54} $\pm$ 2.50 & \textbf{6.94} $\pm$ 2.55\\
         \bottomrule
    \end{tabularx}
\end{table*}

\section{Discussion}

\subsection{DCP Results Analysis}

%C'est trés bien, mais il manque le psotionnement par rapport aux travaux existants. c'est qq chose d'apprécié en general. donc essaye d'inclure qq travaux en plus 3/4 par ex.

First, with higher AP rates, all the fusion models show a better clinical acceptability than the base predictors. This show the overall usefulness of using model fusion algorithms in general. With the EP rates remaining stable, the improvements come mainly from a shift of BE to AP. Among the fusion models, the improvements are much more significant for the DCP models with 10.53\% (DCP 1) and 11.54\% (DCP 2) more AP, demonstrating the superiority of the proposed fusion algorithm.

This improvement is made possible by the increased accuracy in the predicted variations represented by the dRMSE. Nonetheless, we can notice a loss in the prediction accuracy (RMSE). This is an acceptable trade-off since the prediction accuracy remains good enough not to threaten the life of the patient (the clinical acceptability being high).

\begin{figure*}
\centering
\begin{subfigure}[c]{0.1\textwidth}
\caption{ACP}\label{fig:acp_cg-ega}
\end{subfigure}%
\begin{minipage}[c]{0.825\textwidth}
\includegraphics[width=\linewidth]{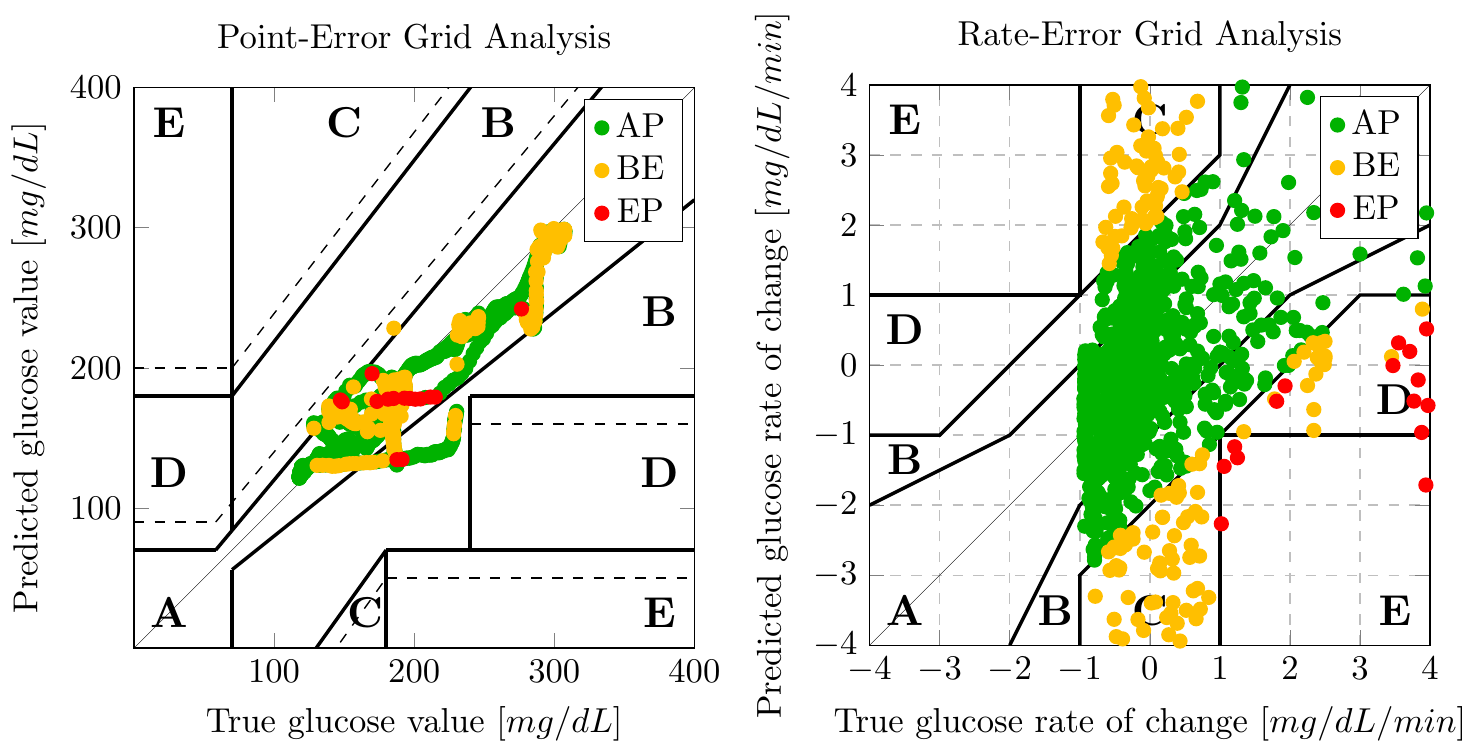}
\end{minipage}
% \\[1ex]
\begin{subfigure}[c]{0.1\textwidth}
\caption{DCP2}\label{fig:dcp2_cg-ega}
\end{subfigure}%
\begin{minipage}[c]{0.825\textwidth}
\includegraphics[width=\linewidth]{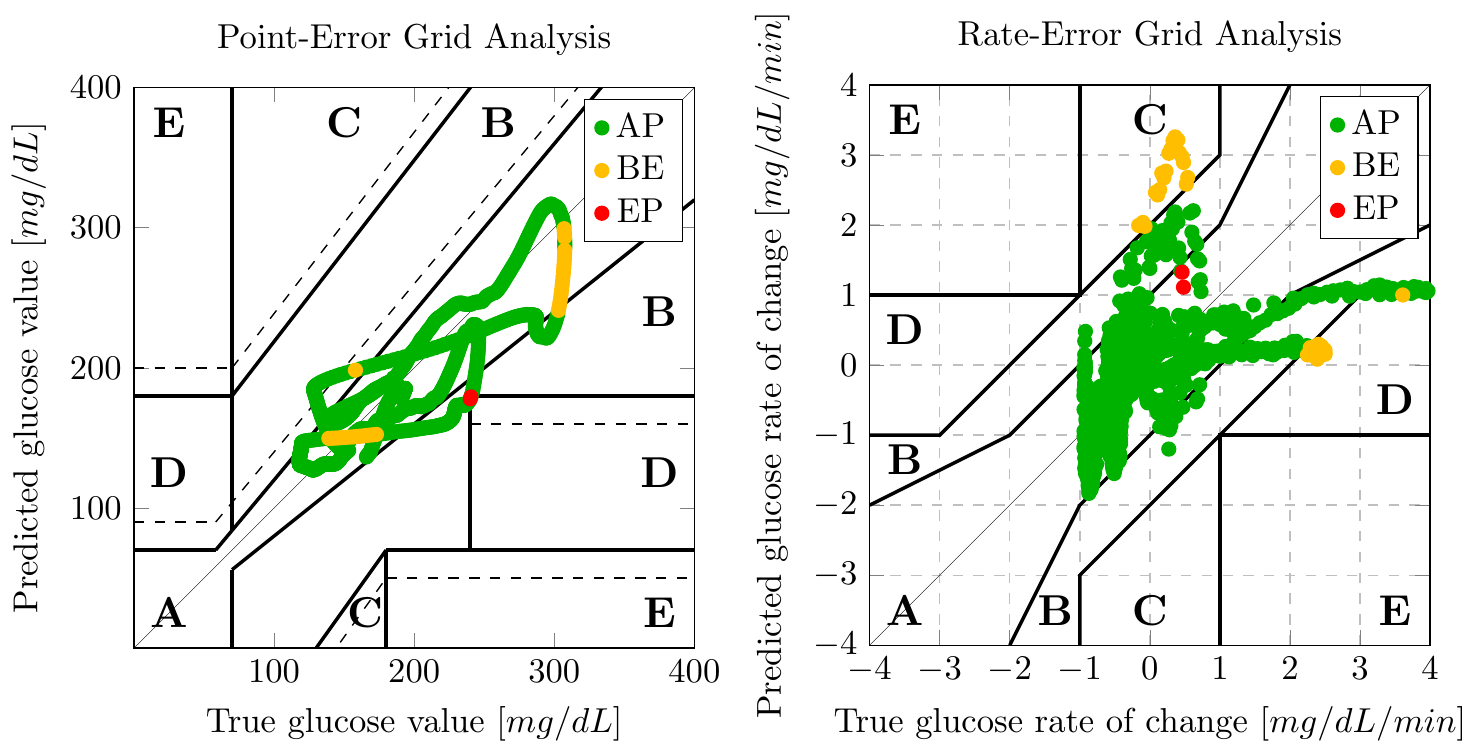}
\end{minipage}
\caption{P/R-EGA of the ACP \subref{fig:acp_cg-ega}, and DCP 2 \subref{fig:dcp2_cg-ega} models for a child during a specific day. Every prediction is assigned a mark from A to E in both grids depending on the ground truth. Then, based on both marks, the prediction is said to be an Accurate Prediction (AP, in green), a Benign Error (BE, in orange), or a Erroneous Prediction (EP, in red).}
\label{fig:cg-ega}
\end{figure*}

Figure \ref{fig:graph_fusion} and \ref{fig:cg-ega}, taken from a particular day of one of the patients, illustrate those dynamics. In a first hand, the P-EGA (representing the accuracy of the predictions) of the DCP 2 model is slighly worse than the one of the ACP model. But, in the other hand, the R-EGA of the ACP model is considerably worse than the one of the DCP 2 model, the predicted variations of the ACP model being more spread out from the optimal diagonal line.

\subsection{Influence of the MSE\textsubscript{DB} loss function}

\begin{figure*}
\centering\begin{adjustbox}{width=0.90\linewidth}
\includegraphics[width=\linewidth]{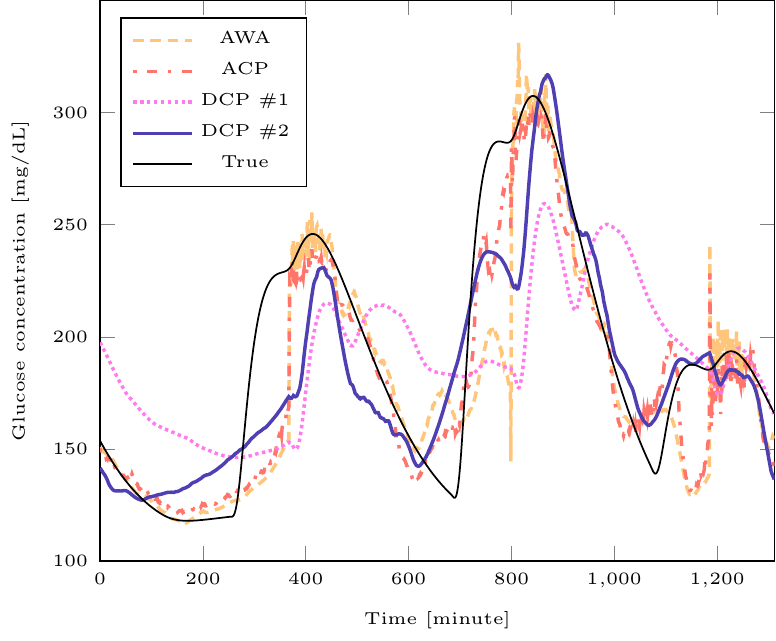}
\end{adjustbox}
\caption{Daily glucose predictions of the fusion models against ground truth for a child during a specific day.}
\label{fig:graph_fusion}
\end{figure*}

Finally, DCP 2, with a lower RMSE and a higher clinical acceptability when compared to DCP 1, shows the importance of using the MSE\textsubscript{DB} loss function. The right side of Figure \ref{fig:graph_fusion} depicts its influence on the training of the dModel: the predicted values are closer to the true values because the model reacts faster to the changes of glucose variations.

% \textcolor{red}{However, the mean accuracy of the predictions, measured by the RMSE, does not see a lot of improvement. It even deteriorates for both DCP models when compared to the best base predictor (FFNN). While this does not seem very good, we have to remember that we are dealing with long-term predictions and we are, therefore, more interested in knowing the general trends of future glucose variations, for which the CG-EGA is a good indicator.}

% \textcolor{red}{Overall the AWA model seems to be the less interesting fusion algorithm in this study. This can be explained by the nature of the fusion. It weighs the base predictors depending on the accuracy of their recent past predictions. However, when making predictions at a PH of 120 minutes, AWA only knows the accuracy of the predictions made up to 120 minutes ago, which are not very representative of the current state of the glucose variations.}

% \textcolor{red}{While both DCP models have worse RMSE values than all the other models, they have a drastically better ESOD (an improvement in the order of magnitude of 3). Those results are due to the inherent nature of the predictions made by DCP algorithm. Because the DCP predicts the glucose derivatives and then sum-up the derivatives to obtain the glucose predictions, predictions errors come from several tiny errors made across time. This makes the predictions themselves at the same time less accurate but also more stable (this behavior is depicted by the right side of Figure \ref{fig:graph}).}

\section{Conclusion}

In this work, we proposed a new fusion algorithm, the Derivatives Combination Predictor, that, by predicting future glucose values through the prediction of its variations, improves the clinical acceptability of long-term glucose predictions. 

To enhance the accuracy of the predictions, we introduced a new loss function, the MSE\textsubscript{DB}, specifically designed for the DCP algorithm.

As a conclusion, the DCP fusion algorithm seems to be promising at addressing the problem of long-term glucose predictions. In future studies, we aim to investigate the use of other models inside the dModel module of the DCP as well as the use of the algorithm itself inside an end-to-end glucose predictive model.

\section*{Acknowledgment}

This  work  is  supported  by  the  "IDI  2017"  project  funded by the IDEX Paris-Saclay, ANR-11-IDEX-0003-02.

\bibliographystyle{IEEEtran.bst}
\bibliography{bibtex.bib}

\end{document}